\author{
\IEEEauthorblockN{Shmeelok Chakraborty}
\IEEEauthorblockA{University of Michigan\\
shmeelok@umich.edu}
\and
\IEEEauthorblockN{Yuewen Hou}
\IEEEauthorblockA{University of Michigan \\
isaachyw@umich.edu}
\and
\IEEEauthorblockN{Ang Chen}
\IEEEauthorblockA{University of Michigan\\
chenang@umich.edu}
\and
\IEEEauthorblockN{Gokul Subramanian Ravi}
\IEEEauthorblockA{University of Michigan\\
gsravi@umich.edu}
}

\documentclass[10pt,letterpaper,compsoc,conference]{iiswc24}

\usepackage{cite}
\usepackage{amsmath,amssymb,amsfonts}
\usepackage{algorithmic}
\usepackage{algorithm}
\usepackage{graphicx}
\usepackage[dvipsnames]{xcolor}
\usepackage[final]{microtype}
\usepackage[italic]{mathastext}
\usepackage{libertine}
\usepackage[T1]{fontenc}
\usepackage{textcomp}
\usepackage[varqu,varl]{zi4}
\usepackage[all]{nowidow}
\usepackage{fancyhdr}
\usepackage{subcaption}
\usepackage{tabularx}
\usepackage{booktabs}
\usepackage{hyperref}

\begin{document}


\title{Empowering the Quantum Cloud User with QRIO}

\maketitle
\pagestyle{plain}

\begin{abstract}
Quantum computing is moving swiftly from theoretical to practical applications, making it crucial to establish a significant quantum advantage. Despite substantial investments, access to quantum devices is still limited, with users facing issues like long wait times and inefficient resource management. Unlike the mature cloud solutions for classical computing, quantum computing lacks effective infrastructure for resource optimization. 

We propose a Quantum Resource Infrastructure Orchestrator (QRIO), a state-of-the-art cloud resource manager built on Kubernetes that is tailored to quantum computing. QRIO seeks to democratize access to quantum devices by providing customizable, user-friendly, open-source resource management. QRIO's design aims to ensure equitable access, optimize resource utilization, and support diverse applications, thereby speeding up innovation and making quantum computing more accessible and efficient to a broader user base. In this paper, we discuss QRIO's various features and evaluate its capability in several representative use cases. QRIO's code and demo video can be found at \url{https://github.com/Rio02coder/QRIO-Components}.
\end{abstract}
\section{Introduction} 

Quantum computing is a disruptive technology ever closer to practical use. In the past decade, quantum devices have seen rapid advancement, transforming from laboratory curiosity to technical reality.
Therefore, establishing practical and substantial ``quantum advantage'' is a high priority. 
To deepen our understanding and for the rapid development of quantum technology, however, it is essential to democratize access to quantum devices to a wide global landscape of users. 
Today, quantum devices are only available with some major players (e.g., IBM, Google, Amazon, Microsoft~\cite{IBMQE,AWS,Azure} and a few others~\cite{IONQ-web,Xanadu-web,Infleqtion-web,Rigetti-web,Quera-web}), and devices are few and far between even among these. This trend is expected to continue until high-fidelity, scalable quantum systems can be developed at a cost that can be amortized. Achieving this depends on researchers and practitioners having enough access to current devices for quick experimentation. Until that point, and probably even afterward, quantum devices and their associated tools will mainly be accessed globally through network access.
Unfortunately, the best way to provide this access to users is not well understood. Currently, each user independently decides how to map their job to a device, resulting in a collectively suboptimal outcome. Real-world measurements have shown that due to inadequate resource management, thousands of quantum jobs are queued on today's devices, with wait times often extending to several days~\cite{Qcloud_Ravi:2021}. This makes accessing quantum resources extremely challenging and significantly hampers our pace of innovation. As the popularity and demand for quantum computing continue to grow~\cite{IBM-users}, we can expect these access issues to become even more pronounced.

\begin{figure}[htbp]
\centering
\includegraphics[width=\columnwidth]{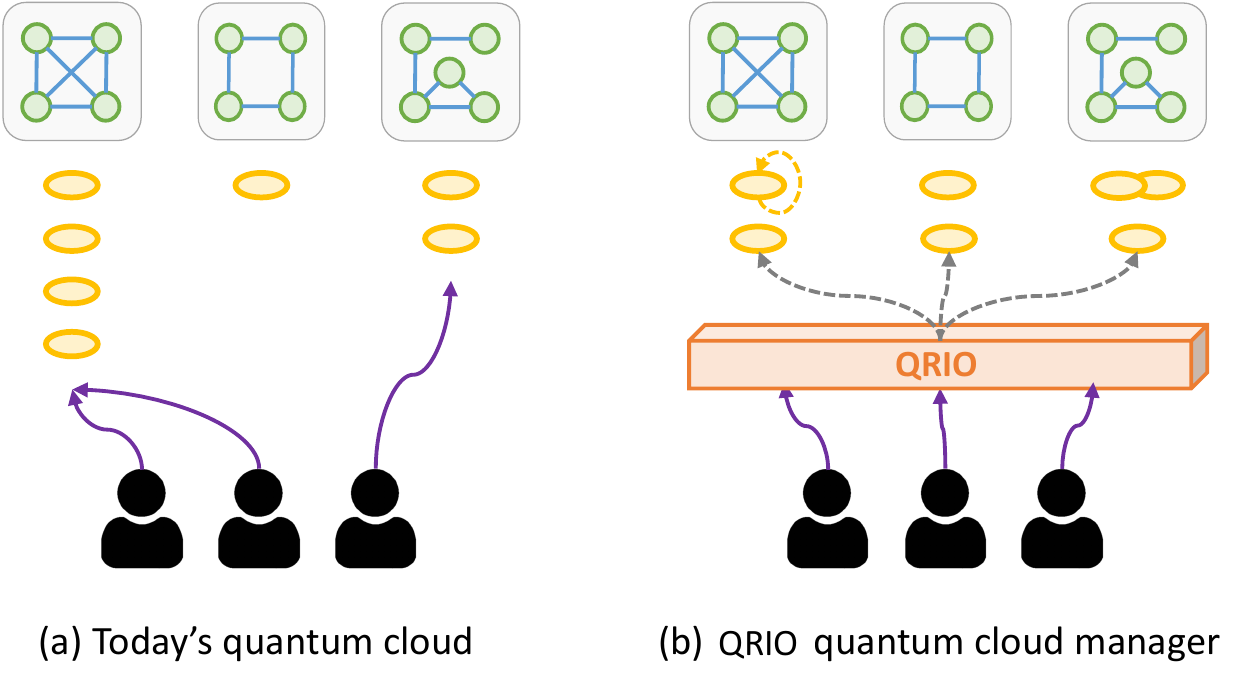}
\caption{Experimenting with Resource Management in the Quantum Cloud with QRIO.} 
\label{fig:overview}
\end{figure}

State-of-the-art cloud computing solutions solve many of these problems for classical computing resources---multiplexing resources across users, no prohibitive setup cost, and high utilization for the cloud providers. Cloud users can spin up compute resources within minutes or seconds, and 94\% of all enterprises use cloud resources of some kind~\cite{cloudwards-report, rightscale-report}.  
We believe that quantum computing could follow an analogous route: 
if we enable a similar paradigm for quantum resource access, this will not only maximize the utility of today's quantum devices but also accelerate innovation for future quantum technologies. 
Although today's quantum devices are already hosted in data centers, and while users submit jobs over the network, the ability to effectively schedule jobs and manage quantum resources, to the benefit of users and vendors, is minimal.
In our envisioned setup, our effective resource management in the cloud will ensure equitable access for all users, allow for varying levels of user control, optimize resource utilization and balance load, and maximize the potential of heterogeneous quantum computing resources for diverse user applications. While relevant to all quantum resources in the cloud, this is especially useful for academic quantum testbeds with quantum computers to outsource but do not possess the infrastructure to manage resources.
In this work, we lay the groundwork toward the goal of building a practical Quantum Resource Infrastructure Orchestrator or \emph{QRIO}, which is illustrated in Fig.\ref{fig:overview}.

In today's era of Noisy Intermediate Scale Quantum (NISQ) Computing and the expected upcoming decade of Early-Stage Fault Tolerance (EFT), we are expected to work with devices that have 100s to 1000s of qubits, with limited qubit connectivity and imperfect operation (quantum gate) characteristics. Concretely, in current setups a general user is faced with three issues: 

\textbf{Usability hurdles}: We lack quantum computers that can produce high-fidelity results for all types of quantum circuits. This is primarily related to the structure of the quantum circuit (more qubits and deeper circuits are harder to run) and the topology and error rates of the quantum device (sparse topology and higher error rates lead to poor execution fidelity). Furthermore, a user's quantum circuit typically requires modifications to match the device topology as well as to avoid highly erroneous qubits, resulting in additional gate operations added to the original circuit. Thus, the decision of choosing the most appropriate quantum computer for a given quantum circuit is a non-trivial task for a general user. This is even more complicated due to the dynamically varying loads on different quantum devices, among other factors. Today, however, users are expected to choose a specific quantum device, which leads to inefficient execution and poor utilization. Thus having an automated framework to manage these complex resources is important to support diverse users and vendor devices with different capabilities.

\textbf{Lack of customization}: While choosing a target device can be challenging for a user, one would imagine that users with reasonable expertise will be aware of certain device characteristics that they desire for their applications---e.g., number of qubits, average gate error, some fidelity expectations, high-level device topology, classical resources, and so on.
As discussed earlier, current quantum clouds do not provide users with the ability to specify just a few characteristics that they desire with which a device can be chosen---users need to choose one specific target device.
Current quantum cloud clusters also do not provide the ability for users to select a set of quantum and classical (e.g. HPC) resources with which they might execute a distributed hybrid quantum-classical application.
The ability to customize resources is critical as we scale up to diverse, novel real-world quantum applications, each of which can have fairly unique requirements for maximum efficiency.

\textbf{Closed Source}: Resource management frameworks employed today are not only lacking in the areas described above, but they are also closed source. Considering that these are today closely tied to specific vendors, it is hardly surprising that they are proprietary. However, building an open-source resource management platform that can be easily accessed, studied, and modified, will accelerate innovation in this space, as has been evident in the classical cloud computing world. It will also provide better interoperability and portability across different current and future quantum computing platforms. As mentioned earlier, this is especially important for academic quantum test-beds that wish to provide quantum resource access to a large set of users but cannot afford state-of-the-art licensed resource management systems. Thus an open-source system is key to rapid progress in quantum cloud resource management.

\noindent\textbf{Contributions:}
To combat these limitations, we develop QRIO which provides a customizable quantum-classical open-source system, addressing the following goals. 
\textit{Ease of use}: QRIO aims to accommodate a wide variety of users of quantum computers, ranging from those who simply have some knowledge of the execution fidelity they desire to more sophisticated users capable of specifying the desired topology and qubit error rates for their circuits.
\textit{Customizability}: QRIO offers the power to the end user to customize their requirements. They can specify their computing needs (both classical and quantum) like CPU, Memory, Qubits, acceptable gate and readout error rates as well as coherence times. Moreover, we allow the user to specify unique quantum device topologies that they desire. Users can also ask for the capability to connect to classical resources. With all this information, QRIO returns the best possible resources to the user to cater optimally to their application.
\textit{Open source}: The codebase of QRIO is open source and will be publicly available on GitHub - \url{https://github.com/Rio02coder/QRIO-Components}. This increases the transparency and interoperability for the user. A demo video is also provided at the above link.

Users can easily interact with  QRIO through a dashboard for a range of experimentation, with the ability to specify any of the following: job circuit (as a QASM file), desired fidelity, desired topology, and desired device characteristics. For all setups, QRIO produces detailed logs and execution results. QRIO is built on top of Kubernetes~\cite{Kub}, which allows us to build our quantum scheduling strategies, support quantum simulated nodes, label nodes with their properties like qubits, CPU, and Memory, and allow communication within and outside the cluster---i.e., the user can add support for hybrid quantum-classical architectures. In terms of supported quantum execution platforms, QRIO allows the user to customize their usage. Users can create clusters comprised of a mixture of noise-free simulators (e.g. QASM simulators), noisy simulators with varying noise models and topologies, different quantum devices, and other classical resources to perform non-quantum tasks.

We demonstrate \emph{QRIO}'s utility via three use-cases:
\begin{enumerate}
    \item \textbf{Filtering quantum resources based on user-specified requirements}: When scheduling a job in QRIO, a user can specify bounds on specific device characteristics (e.g., maximum 2-qubit error rates). QRIO utilizes an in-built filtering mechanism to filter out devices that do not match user requirements and only runs resource management, scheduling, and transpilation tasks on shortlisted devices. As device size, heterogeneity, and technology options grow in the coming years, such filtering schemes will considerably reduce classical pre-processing overheads while enabling optimal resource allocation.
    
    \item \textbf{Fidelity-requirement based resource allocation}: In scenarios wherein a user has a rough estimate of the execution fidelity requirement for their target applications, the user can specify this requirement and QRIO can allocate resources that loosely match the user's fidelity requirement. This is achieved via the use of Clifford circuits~\cite{gottesman1998heisenberg} and scalable classical simulation. As circuit complexity continues to increase, simplistic analytical methods of fidelity estimation fail, whereas Clifford-based methods have shown high accuracy. Thus, QRIO's use of a Clifford-based method will enable long-term automated resource allocation and scheduling with high accuracy. 
    
    \item \textbf{Topology-requirement based resource allocation}: In scenarios wherein a user knows the hardware topology best suited for their quantum task (easily discernible for optimization problems, for example), the user can specify their topology requirement and QRIO can allocate resources that loosely match this requirement. This is achieved via the use of subgraph isomorphism-based techniques. While currently leveraging some aspects of the state-of-the-art Mapomatic~\cite{Nation_2023} approach, the long-term vision is for a scalable methodology that can handle many 1000s of qubits. Thus, QRIO's topology-based allocation allows for quantum resource assignments that are tailored to the structure of the target user application.
\end{enumerate}

\section{Background and Related Work}

\subsection{Qubits and Quantum Gates} 

The quantum bit, or qubit, is the basic computational unit of quantum computing. A qubit can superimpose both states, in contrast to a traditional bit, which can only be either 0 or 1. $|\psi\rangle = \alpha|0\rangle + \beta|1\rangle$ represents the most general state of a qubit, where $\alpha$ and $\beta$ are complex numbers that fulfill the equation $|\alpha|^2 + |\beta|^2 = 1$. Measurements collapse the qubit from a superposition to either 0 or 1. Quantum gates utilize qubits to execute quantum calculations.
Quantum gates and measurement operations are noisy in the present Noisy Intermediate-Scale Quantum (NISQ) era computers; two-qubit operations are especially noisy compared to single-qubit operations. 

\subsection{Quantum Heterogeneity}
Quantum computing is not limited to a single technology; instead, it encompasses various modalities based on superconducting transmons, trapped ions, neutral atoms, photons, and more. Prior work~\cite{Supermarq} discusses in detail that different devices can differ significantly in terms of the number of qubits,  qubit coherence times, quantum gates times, different types of error rates, device topology, and more. Furthermore, there is substantial spatial and temporal variability in qubit error characteristics within devices. Spatial variability refers to the different error rates and gate properties for different qubits within a device. ~\cite{Qcloud_Ravi:2021} has shown more than an order of magnitude difference in characteristics like two-qubit gate error and readout error within IBMQ devices. Temporal variability refers to the changing properties of qubits over time---~\cite{Qcloud_Ravi:2021} has shown 2-3x variation in 2-qubit gate characteristics across calibration cycles (each calibration cycle roughly corresponding to one day). From the above discussion, it is intuitive that the same application executed on different machines or on different qubits within a machine or at different times, can experience different noise characteristics and therefore impact execution differently. This makes optimal resource management an especially vital but challenging task for quantum computing in the cloud.

\subsection{Quantum Transpilation}
The compilation of a quantum program is similar to that of a classical program. The transpiler processes a quantum circuit as input and performs a series of transformations (known as transpiler passes) to generate an executable that adheres to the constraints of a particular device (such as qubit connectivity and native gate set) and aims to minimize the anticipated error. One of the most well-known quantum transpilers, the IBM Qiskit transpiler, achieves this through six broad steps \cite{Qiskit}: Virtual Circuit Optimization, 3+ Qubit Gate Decomposition,  Placement on Physical Qubits, Routing on Restricted Topology, Translation to Basis Gates, and Physical Circuit Optimization; this is broadly representative of other quantum transpilers as well. 

\subsection{Classical Cloud Resource Management}

Job scheduling and cluster management are essential and have been prevalent since the days of distributed computing. In this section, we discuss relevant background in the context of large-scale cluster schedulers and management systems both in the classical and the quantum domain.

In most High Performance Computing (HPC) clusters, resource schedulers are used to manage resources and ensure their fairness. However, most of these schedulers are either closed-source or have limited customizability and thus cannot be modified for QRIO. \textit{SLURM}\cite{inproceedings} is an open-source workload manager and scheduler designed for high-performance computing environments. It is widely used in academic and research institutions for scheduling large-scale computing jobs. The primary limitation of SLURM is that its scheduler is mainly optimized for batch jobs in an HPC environment. Moreover, the kind of workloads run on QRIO is containerized. So, each quantum job run in the cluster is given its own environment and does not communicate with other jobs. The workloads that SLURM manages and schedules favor a batched workload where resources are shared among the jobs.

\textit{Google's Borg} \cite{43438} offers a cluster management system with an inbuilt scheduler that is capable of running thousands of jobs from various sources across multiple clusters, each comprising hundreds to thousands of machines. It is capable of scaling with high utilization, leveraging techniques like admission control, machine sharing, and process level isolation. The reason QRIO cannot be built upon Borg is that it does not offer an open-source implementation. It is meant strictly for scheduling and managing Google's workloads. This implies that Borg lacks the portability and flexibility to adapt to changing needs. Moreover, the components of Borg are not modular enough to allow customizability QRIO needs to utilize a classical scheduler as a scheduler for quantum workloads.

\subsection{Prior Work on Quantum Job Scheduling}
Ravi et al.~\cite{9605297} proposed an efficient job scheduling mechanism factoring in aspects like queuing times, fidelity across machines, and user quality of service requirements. However, this system utilizes a simplistic prediction function for the scheduling of jobs and does not interface with any state-of-the-art cloud management infrastructure. QRIO, on the other hand, is built on Kubernetes, meaning that it can be directly deployed in the cloud with minimal modification to manage quantum devices. QRIO also provides users with multiple options with which they can specify resource requirements and a fully functional end-to-end framework. 
\section{Design and Implementation}

\begin{figure*}[htbp]
\centering
\includegraphics[width=0.75\textwidth]{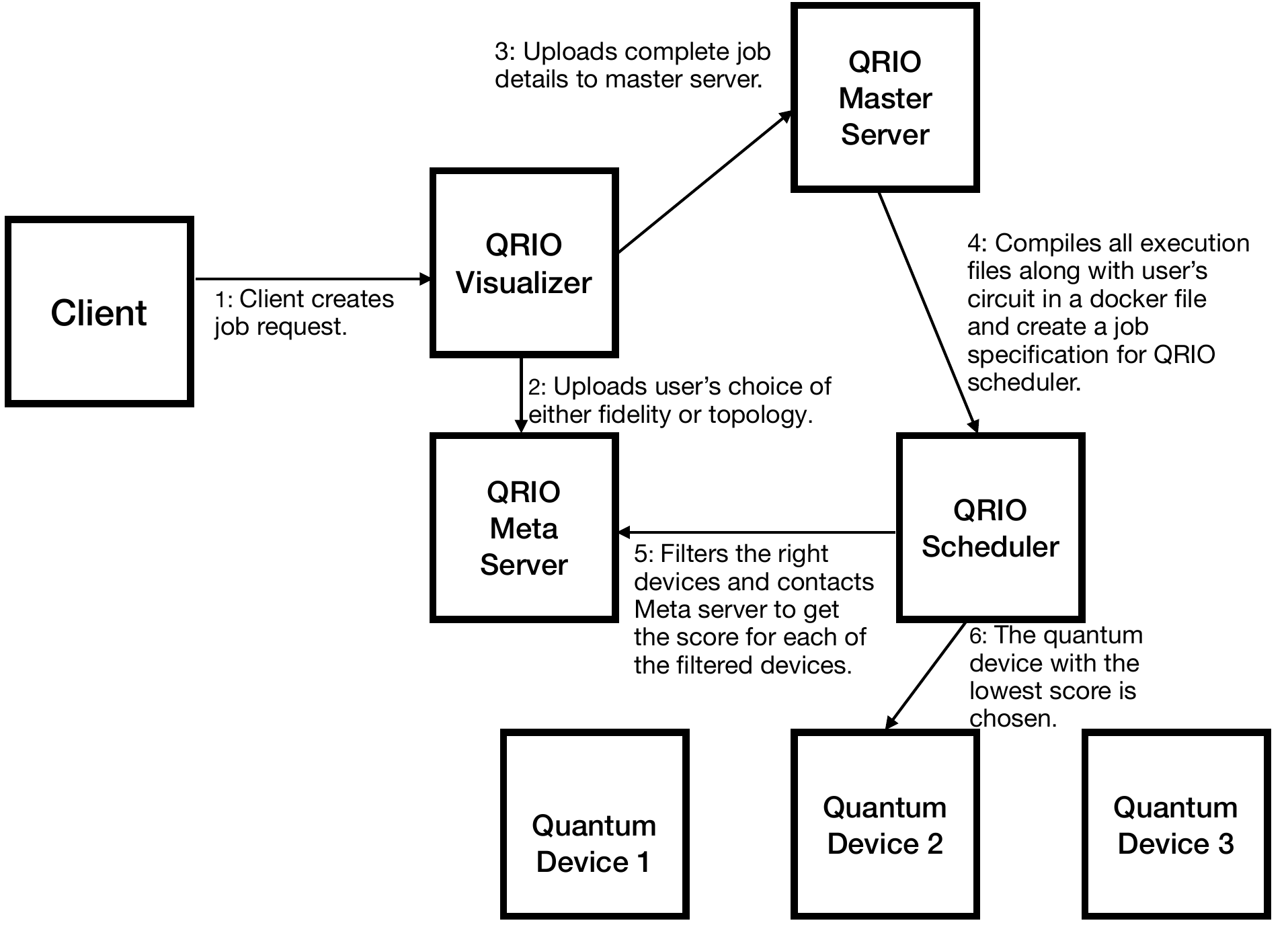}
\caption{System design for QRIO cluster of 4 nodes} 
\label{fig:QRIO System Design}
\end{figure*}

QRIO provides a complete user-centric infrastructure, which is geared towards automatically scheduling quantum jobs to a cluster of quantum nodes. The infrastructure gives the user a range of specifications, which could be as simple as a fidelity threshold to custom topology requirements for their jobs. To support this infrastructure, QRIO is comprised of four key components---\textbf{QRIO Visualizer}, \textbf{QRIO Master Server}, \textbf{QRIO Meta Server} and \textbf{QRIO Scheduler}, as shown in figure \ref{fig:QRIO System Design}.
Roughly a client using QRIO goes through the following cycle for scheduling their quantum job.

\begin{itemize}
    \item The client uses the \textbf{QRIO visualizer} to create a `job' by submitting the quantum circuit(s) along with details such as the number of qubits, topology, and fidelity requirements.
    \item \textbf{QRIO Visualiser} then contacts the \textbf{QRIO Meta Server} with the details on whether the user chooses a desired fidelity or a desired topology and also uploads relevant files.
    \item \textbf{QRIO Visualiser} then sends the complete job request containing all details to the \textbf{QRIO Master Server}.
    \item \textbf{QRIO Master Server} then compiles the job circuit along with some other files required for execution into a docker file which is uploaded to a docker hub. It also creates a Job specification file for the scheduler to read. Once the specification is ready it is sent to the \textbf{QRIO Scheduler}.
    \item \textbf{QRIO Scheduler} first reads the specification and filters the appropriate quantum devices for the job. For each of these devices, it contacts the \textbf{QRIO Meta Server} for a score. The \textbf{QRIO Meta Server} uses the user-supplied information about the desired topology or desired fidelity to return a score.
    \item \textbf{QRIO Scheduler} upon receiving a score for appropriate devices chooses the device which has the lowest score for scheduling.
\end{itemize}
In the following sections, we describe our cluster setup and the 4 key components in detail.

\subsection{QRIO Cluster} \label{Kubernetes Cluster} 

The QRIO cluster is at the heart of QRIO. It provides nodes that each integrate classical and quantum resources (although we use quantum simulators in our experiments). We choose to build this cluster over Kubernetes due to its flexibility and modularity. In the Kubernetes architecture, individual components like Nodes, Jobs and Scheduler can be customized to fit the user's needs. Moreover, using Kubernetes gives the desirable properties of distributed systems, such as availability and horizontal scalability. QRIO can thus self-restart nodes and jobs if they are down, and can also horizontally scale to a large number of nodes, provided the computing capability (i.e. CPU and RAM to host such a cluster) is available.

A QRIO cluster is generally defined by the vendor with a specific number of devices, each with unique characteristics. A cluster traditionally contains a master node and a set of worker nodes as shown in \ref{fig:QRIO System Design} (the master node contains QRIO Scheduler and worker nodes are Quantum devices 1 to 3). The master node is where the scheduler is run and is responsible for managing other worker nodes (which house the quantum+classical compute).

Each worker node in the cluster is set up with a backend file written in Python by the vendor. This backend file carries information from the node's quantum device (real or simulated)~\cite{qiskit23}. Specifically, this backend object requires the vendor to provide at least the following information 
\begin{itemize}
    \item Coupling map for the device.
    \item Two qubit error rates for the device.
    \item Single qubit error rates for the device.
    \item Readout error rates for each qubit in the device.
    \item Readout length for each qubit in the device.
    \item T1 and T2 times for each device.
    \item Basis gates for the device.
\end{itemize}

The above parameters are the mandatory information in the current QRIO version. However, we do not restrict the vendor to just these parameters. The vendor can provide further details about the device like pulse characteristics. When the vendor has finalized this backend file, a copy of it is also kept in the \textbf{QRIO Meta Server}.
As a standard, all nodes are required to contain this file named \textit{backend. py}, and the Qiskit Backend object is exposed in a variable called \textit{backend}. When a job is scheduled on a node, it reads the \textit{backend} object from its \textit{backend.py} file and uses it as the quantum device running their quantum job (more details on this functionality have been provided in \ref{QRIO Master Server}).

Lastly, we label each node in the cluster with its properties which helps Kubernetes in the scheduling process of a job (more details
in \ref{Kubernetes Scheduler}). Concretely, we specify the following parameters :
\begin{itemize}
    \item Number of qubits.
    \item Average two-qubit gate error.
    \item Average T1 and T2 times for the entire device.
    \item Average readout error rate.
    \item CPU and Memory capacity of the node.
\end{itemize}
When scheduling a job to the cluster, the user can specify some or all of the above parameters, and devices are selected/filtered based on this information. 

\subsection{QRIO Visualizer} \label{QRIOVisualizer}

\begin{figure}[htbp]
\centering
\includegraphics[width=\columnwidth]{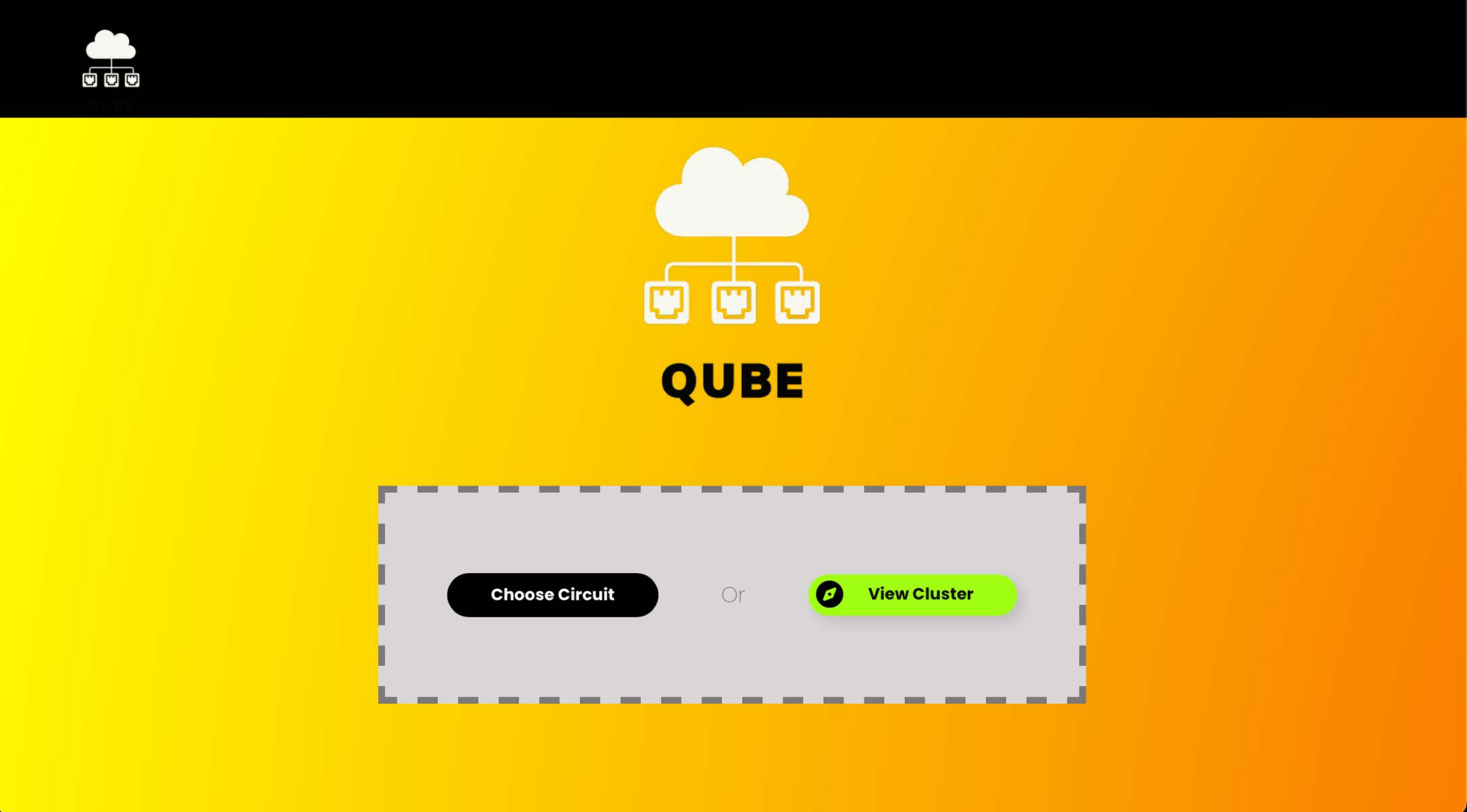}
\caption{QRIO Front Page}
\label{fig:QRIO_Front_Page}
\end{figure}

The QRIO Visualizer is a web application (developed using \textbf{React}~\cite{react_website}) that a client interacts with to submit and check their job logs. The user is first presented with a screen to either choose a circuit or view the current cluster as shown in figure \ref{fig:QRIO_Front_Page}. 

\begin{figure*}[htbp]
\centering
\begin{subfigure}{0.48\textwidth}
  \centering
  \includegraphics[width=\textwidth]{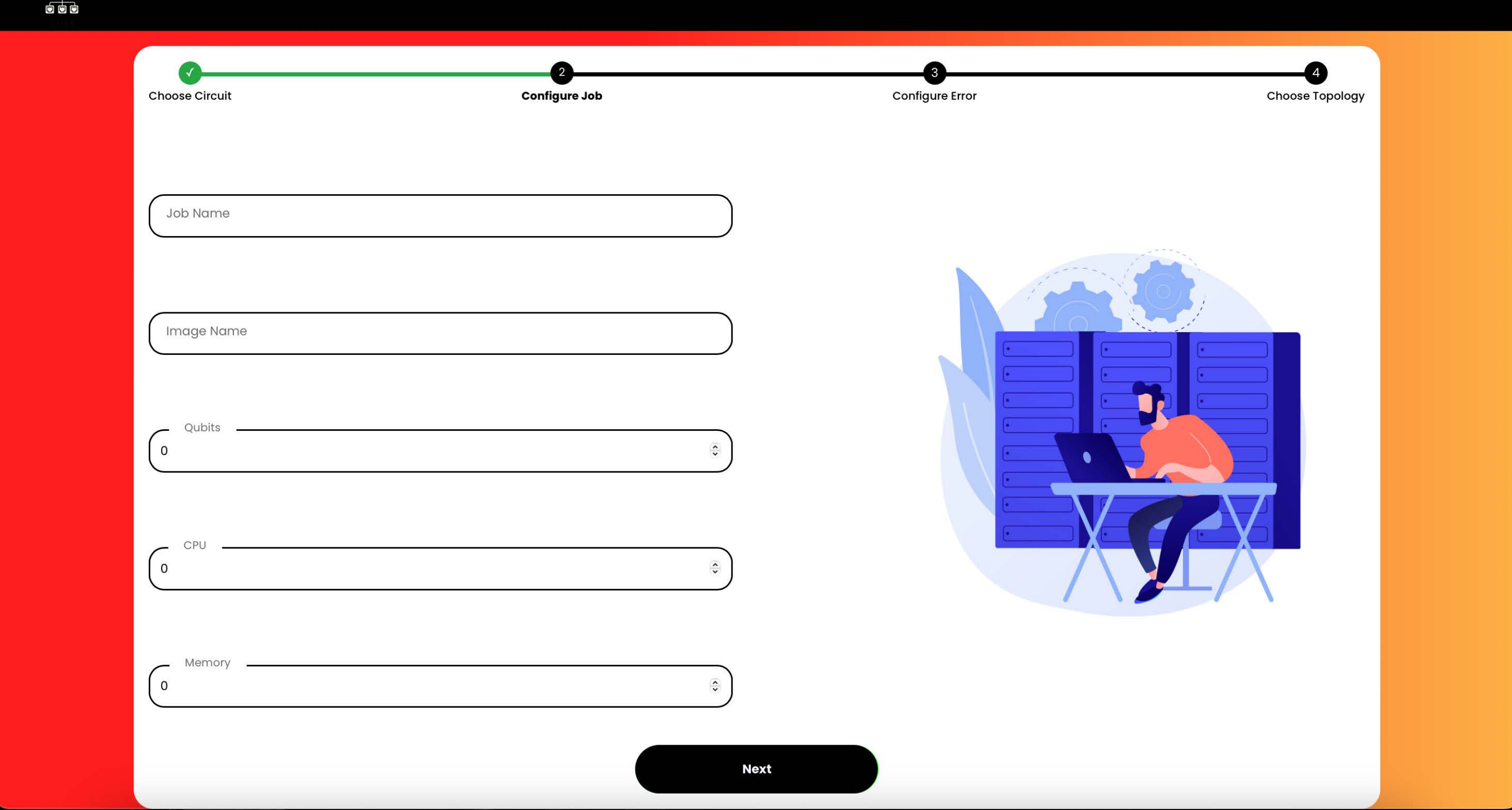}
  \caption{Job input details}
  \label{fig:QRIO_Job_Details}
\end{subfigure}
\hfill
\begin{subfigure}{0.48\textwidth}
  \centering
  \includegraphics[width=\textwidth]{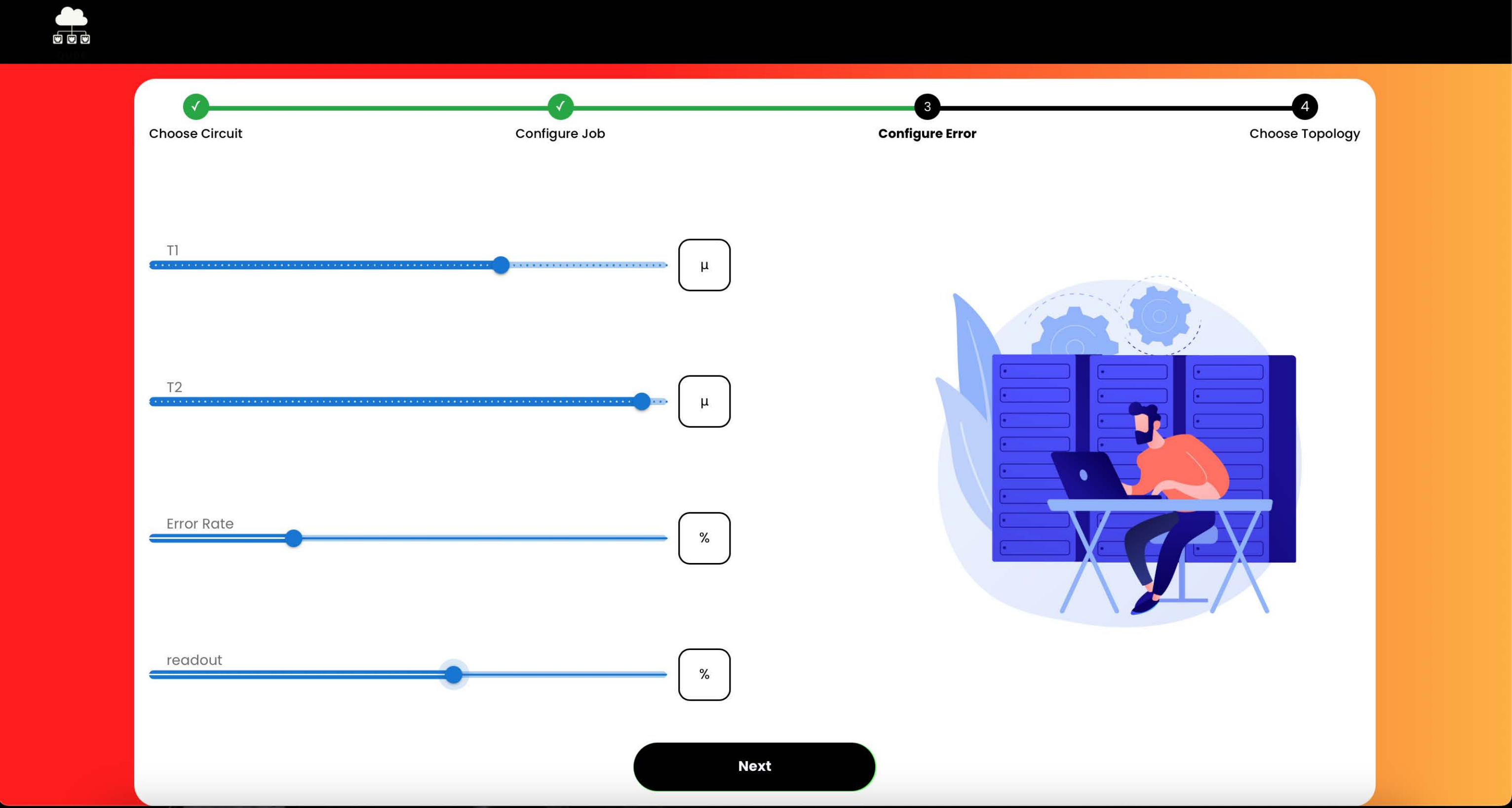}
  \caption{Requested resource characteristics}
  \label{fig:QRIO_Job_Characteristics}
\end{subfigure}
\hfill
\begin{subfigure}{0.48\textwidth}
  \centering
  \includegraphics[width=\textwidth]{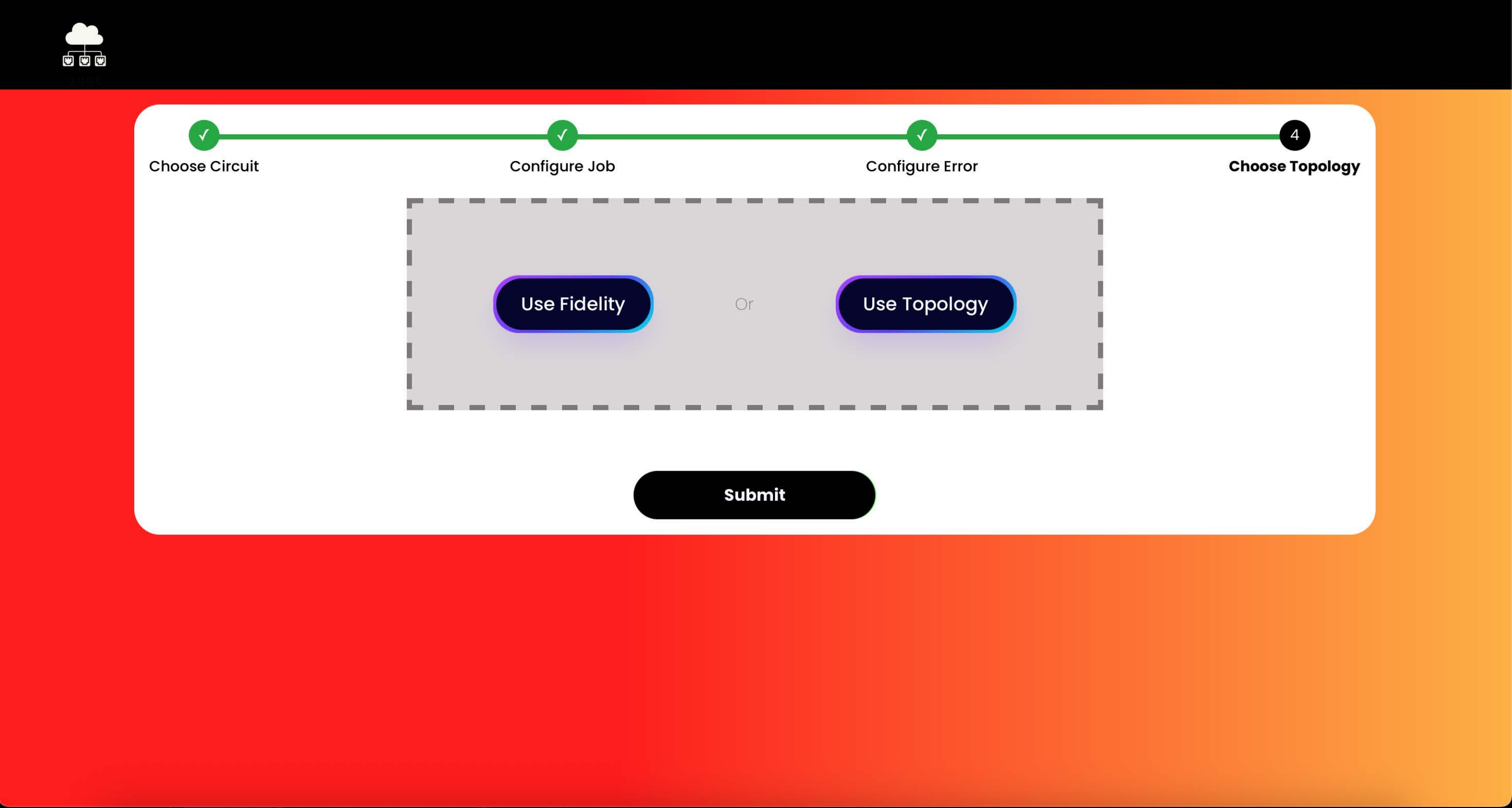}
  \caption{Device select strategy: topology/fidelity}
  \label{fig:QRIO_Topology_or_Fidelity}
\end{subfigure}
\hfill
\begin{subfigure}{0.48\textwidth}
  \centering
  \includegraphics[width=\textwidth]{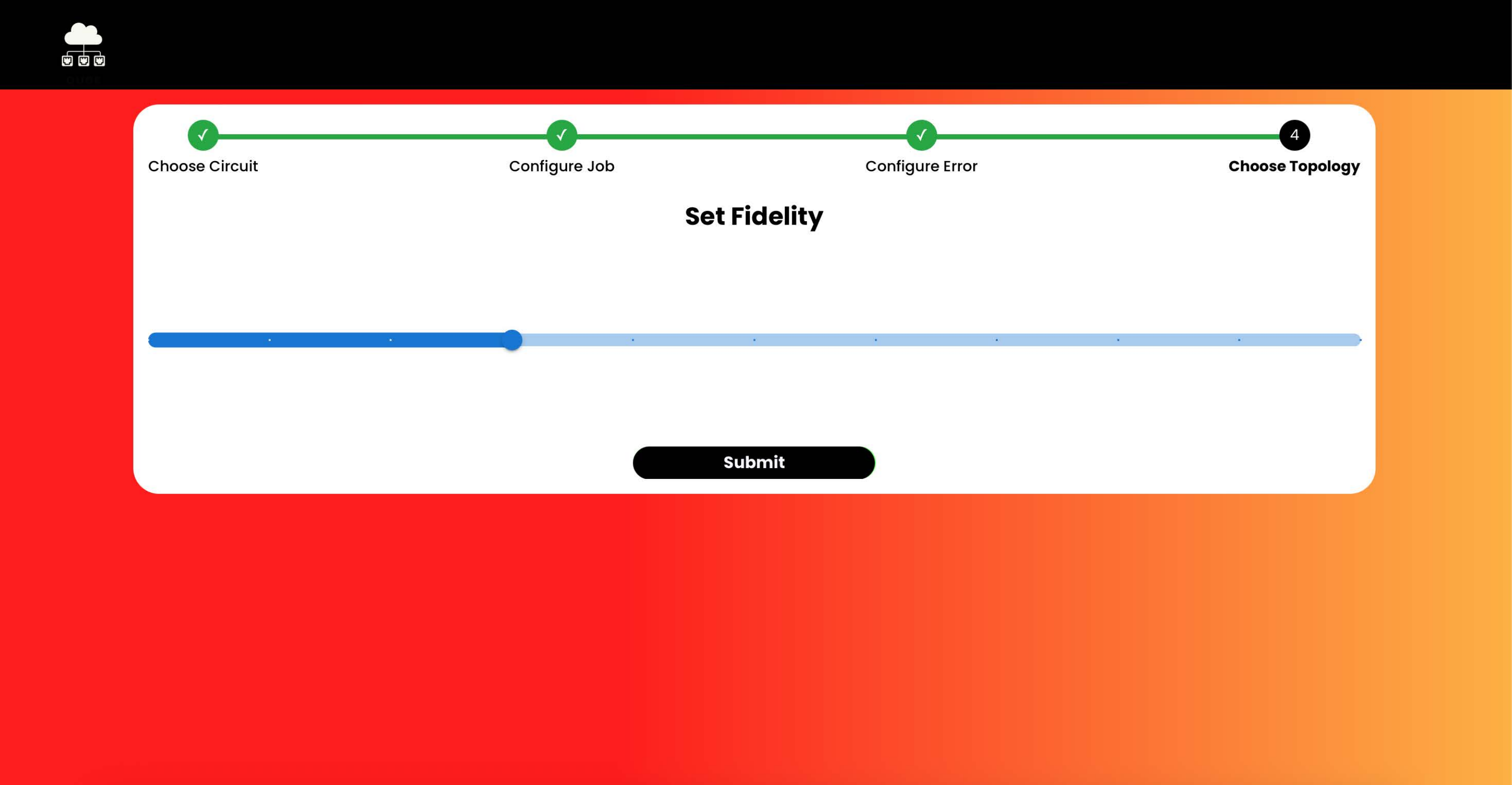}
  \caption{Requested fidelity input}
  \label{fig:QRIO_Fidelity_Form}
\end{subfigure}
\hfill
\begin{subfigure}{0.48\textwidth}
  \centering
  \includegraphics[width=\textwidth]{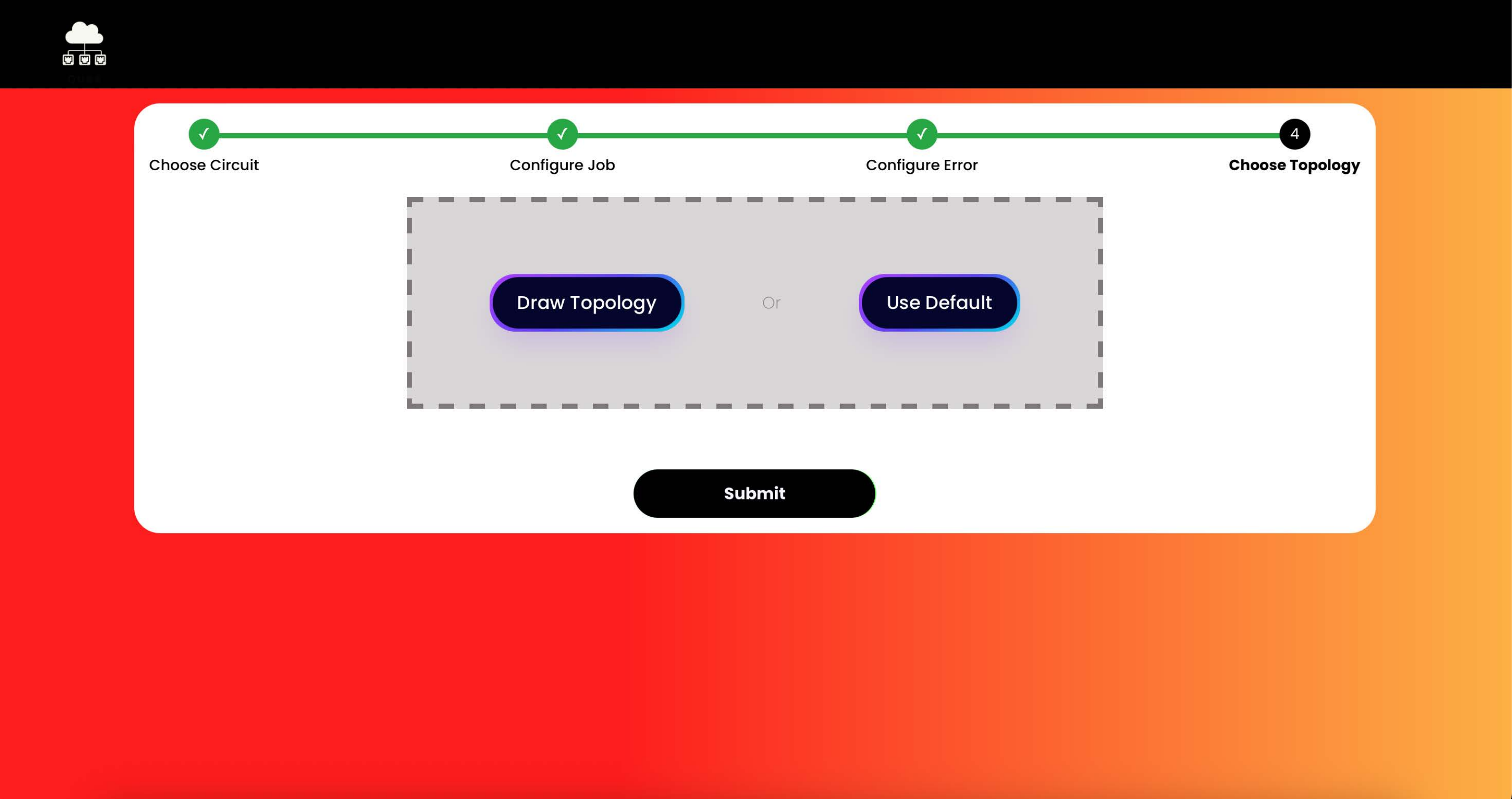}
  \caption{Choosing from default topology}
  \label{fig:QRIO_Topology_Options}
\end{subfigure}
\hfill
\begin{subfigure}{0.48\textwidth}
  \centering
  \includegraphics[width=\textwidth]{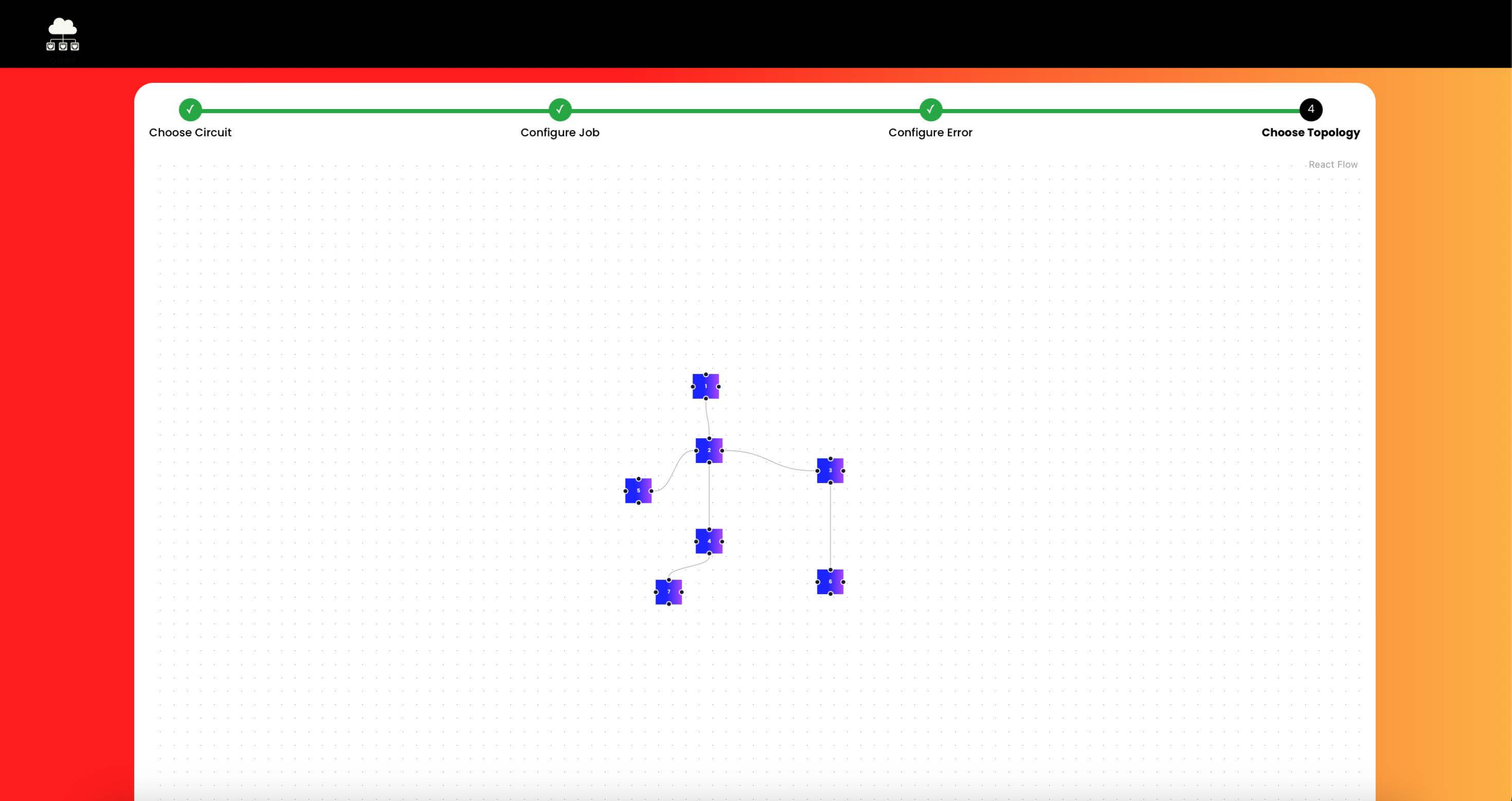}
  \caption{User-created topology request}
  \label{fig:QRIO_Draw_Topology_Form}
\end{subfigure}
\caption{QRIO User Inputs
}
\label{fig:QRIO_Forms_and_Logs}
\end{figure*}

To run a quantum job, the user first needs to choose a circuit represented as a QASM (Quantum Assembly) file. Following that, the user is greeted with a 3-step form to specify details about their job and desired properties of the node where the job is scheduled on. The user has to specify details such as \textbf{Job Name}, \textbf{Docker Image Name}, \textbf{Number of Qubits}, \textbf{CPU requirement}, \textbf{Memory requirement} (in MB) as shown in figure \ref{fig:QRIO_Job_Details}. 
Next, the user has to specify (any) preferred device characteristics. Currently, they can specify T1, T2, average gate error rate, and average readout rate, as shown in figure \ref{fig:QRIO_Job_Characteristics}. In the last step, the user has two options: specify a \textbf{fidelity} requirement or specify a \textbf{topology} requirement as shown in figure \ref{fig:QRIO_Topology_or_Fidelity}. These parameters are integral to the system as they help the scheduler choose the appropriate device for the job. In the former case of choosing a fidelity, the user simply provides a number between 0 and 1 as shown in figure \ref{fig:QRIO_Fidelity_Form}. In the latter case, the user is presented with two options---using one of the default topology options or creating their own required topology as shown in figure \ref{fig:QRIO_Topology_Options}. If creating their topology, the user is presented with a canvas (built over react-flow) containing the specified number of qubits and then the user can draw edges between qubits to specify interactions shown in figure \ref{fig:QRIO_Draw_Topology_Form}. Intuitively, this topology specified by the user embodies the connectivity of a device they consider suitable for the job (and would likely be set to match the connectivity that their application/circuit would ideally prefer). Once the final topology is set, this is converted into a pseudo quantum circuit (more details in Section \ref{topRS}) called a topology circuit. Specifically, this topology circuit is a quantum circuit of the specified number of qubits created and each interaction between two qubits is modeled as a 2-qubit CNOT gate.

Once the workflow completes, the job metadata is sent to the QRIO Meta server, and other details like the number of qubits, Job name, and other requirements are sent to the QRIO master server. The metadata for the job depends on what the user selected in the final step of the form. If the user selects the fidelity option, then the Job name, Circuit QASM file, and the fidelity number are sent over to the server. In the other case of selecting a topology, the created topology circuit file, and Job name are sent to the server (see table \ref{Details to Metaserver Table}). The QRIO meta server keeps track of these and becomes essential while scheduling the job. \par
\begin{figure}[htbp]
\centering
\includegraphics[width=\columnwidth, trim={0 0 0 6cm},clip]{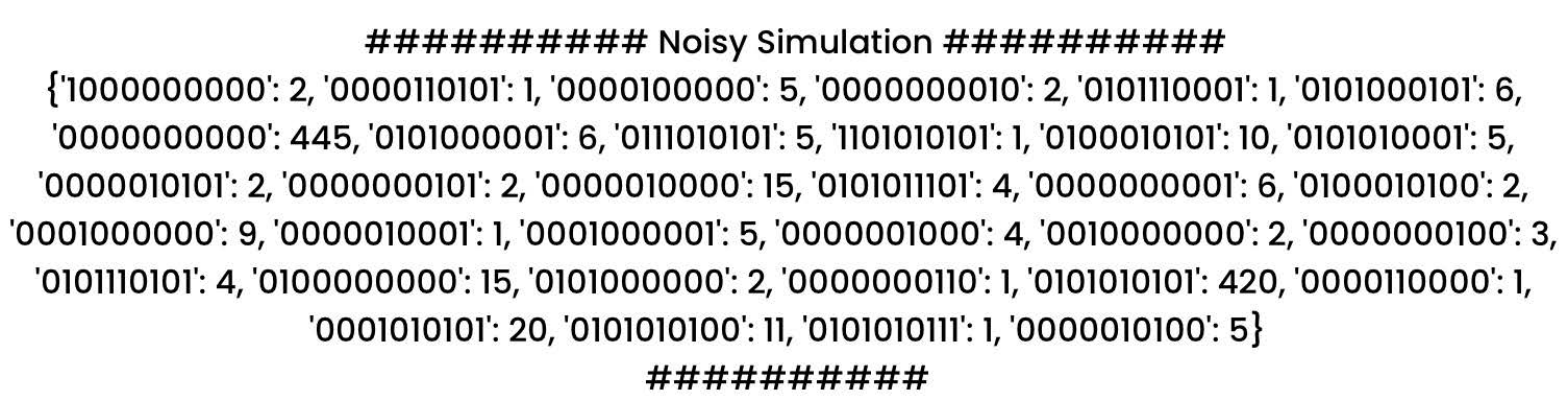}
\caption{QRIO Logs for 10 qubit Bernstien-Vazirani Circuit}
\label{fig:QRIO Logs for Bernstien Vazirani Circuit}
\end{figure}

Lastly, once the job is scheduled the user is presented with a screen where they can check logs and the device their job is scheduled on as shown in figure \ref{fig:QRIO Logs for Bernstien Vazirani Circuit}. When the user clicks on the check logs button, the visualizer contacts the QRIO master server and it returns the logs for the job by contacting the QRIO master node in the cluster. These logs are, however, only available once the job has finished execution, and thus users might have to wait before they see logs.

\begin{table}[htbp] 
\centering
\begin{tabularx}{\columnwidth}{|>{\raggedright}X|>{\raggedright\arraybackslash}X|}
\hline
\textbf{User Chosen Option} & \textbf{Details sent to QRIO Meta server} \\ \hline
Fidelity & Fidelity number, Job Name, Original user circuit as QASM file\\ \hline
Topology & Job Name, User-created topology file \\ \hline
\end{tabularx}
\caption{Details sent to QRIO Meta Server based on the option chosen by the user}
\label{Details to Metaserver Table}
\end{table}

\subsection{QRIO Master Server} \label{QRIO Master Server} 

The QRIO master server is responsible for containerizing the job into a docker file and contacting the QRIO master node in the cluster for scheduling the job. 
As part of the containerization process, the server first creates a python script. This script is designed to read the backend information from the device this job is scheduled to(i.e. the backend file as explained in \ref{Kubernetes Cluster}), transpile the QASM file (user-provided circuit) to the backend, and finally run the transpiled circuit. Since a docker container is an independent running environment, it needs to be populated with adequate packages to run the circuit. Given that our current backends (quantum devices in the cluster nodes) interface with Qiskit, we install the following packages (specified in a requirements file) - qiskit, qiskit-aer, matplotlib, qiskit\_ibmq\_provider, qiskit\_ibm\_runtime \cite{IBMQR}. Finally, the master server takes all these files and creates a directory for the job in the master server's file system where it also writes a docker file for containerization. Concretely, the directory contains:
\begin{itemize}
    \item The user's circuit as a QASM file.
    \item The generated Python script for running the QASM file on the node it is scheduled to.
    \item The requirements.txt file.
    \item Docker file containing the commands to create a docker container having the files in this directory.
\end{itemize} 

The master server then builds the docker file under the specified image name and then pushes it to the docker hub. To schedule a job in QRIO, one also requires to create a Yaml file representing the Job requirements and image name for the docker container of the job. Recall, that for a given job, the QRIO visualizer provides the master server with the Job name, Image name, Number of qubits, CPU and Memory requirement, T1, T2 times, Error rates, and Readout rates. So, the master server constructs the Job Yaml file with the properties passed to it and invokes the QRIO master node to schedule that job.

\subsection{QRIO Meta Server} \label{QRIO Meta Server}
The QRIO meta server is primarily responsible for storing metadata for a job and responding to score requests for the job. Overall, the scoring process for the job is performed concerning a specific backend (quantum device). So, the QRIO meta server stores each backend present inside each node in the cluster (this refers to the backend files defined by the vendor in section \ref{Kubernetes Cluster}). When a job scoring request comes in, it has a request body containing the job name and the device to score it against. On receiving such a request, the server first retrieves the backend object for that device and checks the database if a fidelity threshold exists for the job. If so, that job is scored using a \textbf{Fidelity Ranking} strategy, and if not it is scored using a \textbf{Topology Ranking} strategy. We discuss these in detail next.

\subsubsection{Fidelity Ranking Strategy}
In this workflow, the user specifies an estimate of the fidelity required for their input quantum circuits. Deriving the fidelity requirements of quantum circuits is itself a non-trivial problem, however, we assume that users with reasonable expertise can estimate this for their target applications. For example, a quantum chemistry problem could derive the accuracy requirements of its quantum circuits with the help of the chemical accuracy target and the depth of the quantum circuits in use. However, a user-provided fidelity requirement alone is insufficient---this is because if the user quantum circuit is executed on any quantum device, the fidelity of the output is unknown. After all, the correct noise-free output of the circuit is unknown (because the circuit, if of any practical size, cannot be classically simulated). Thus, we employ Clifford canary circuits as proposed in prior work~\cite{Quancorde_Ravi2022,dangwal2023clifford} which is explained next.

Simulating quantum problems on classical computers generally demands exponential resources. However, there is an exception: the classical simulation of the Clifford space, a subset of the entire quantum Hilbert space. According to the Gottesman-Knill theorem, circuits composed solely of Clifford operations can be precisely simulated in polynomial time, which is almost linear relative to the number of gates and qubits~\cite{gottesman1998heisenberg}. 

Thus, we build Clifford `canary' circuits that are Clifford versions of the user's circuit (i.e. the original circuit without its non-clifford gates). Not only are these circuits classically simulable even for a large number of qubits, but they are also highly representative of the behavior of the original circuit on the quantum device in terms of the effects of noise because they retain the circuit structure (especially the noisy 2-qubit gates) from the original circuit as all of those are clifford operations. These Clifford canary circuits are run both on the device and in classical simulation, and the canary fidelity is calculated on all the devices (this is a shortlisted set of devices based on the filtering process discussed in Section \ref{Kubernetes Scheduler}). Due to the structural similarity between the canary and the user's circuit, the fidelity of the canary is loosely representative of the fidelity of the user's circuit on any given device. Thus, based on the canary fidelity estimates, we select the device that most optimally matches the user's requirements. Note that the fidelity estimate produced by the canary circuit on each quantum device is unlikely to be an exact match for the fidelity produced by the original circuit on those devices. Since this is not a guarantee, a user could provide fidelity requirements on the higher side so that appropriate resources are allocated with higher confidence.

\subsubsection{Topology Ranking Strategy}
\label{topRS}
In this workflow, the QRIO meta server is only provided with the topology file (which was set by the user and sent by the Visualizer) and the devices against which the scoring is being performed. To perform the scoring we leverage features of a library called \textit{Mapomatic} \cite{Nation_2023} which is briefly explained below.

Mapomatic optimizes quantum circuits after compilation to align them with low-noise subgraphs of the quantum device. 
The initial compilation can be performed with state-of-the-art techniques such as SABRE~\cite{DBLP:journals/corr/abs-1809-02573}.
Broadly a two-step process is employed by Mapomatic. Firstly, device subgraphs are identified by traversing the device topology and outlining areas of the devices that are the best fit for the qubit circuit. Secondly, each identified subgraph is scored using a cost function that incorporates device error characteristics to estimate the amount of error the circuit might suffer if it is mapped to that particular subgraph. 
Finally, the subgraph for which the score is the lowest is considered the most suitable location for the target quantum circuit.

In QRIO, recall that the task is to find the device that has a topology that is most similar to the user's specified requirements. So QRIO leverages the topology circuit that was created earlier (Section \ref{QRIOVisualizer}) and then runs Mapomatic on this circuit and all the shortlisted quantum devices (as before, this is a set of devices based on the filtering process discussed in Section \ref{Kubernetes Scheduler}). The Mapomatic-generated best-fit device for this topology circuit is, in essence, the device that most resembles the user's topology requirements. Thus, this device is selected for the user's circuit execution.

\subsection{QRIO Scheduler} \label{Kubernetes Scheduler}
The QRIO scheduler is the primary component used for scheduling a quantum job in QRIO. The entire workflow of the scheduler is broken into many parts, but the two primary stages are - \textit{Filtering} and \textit{Ranking}. In the \textit{Filtering} stage, the scheduler checks which nodes are fit for scheduling,  i.e. which devices meet the resource requirements of the job. In our case, the properties labeled for each node in the cluster and the user-desired properties passed in the Yaml file for the job are compared in the filtering phase to determine which nodes are fit to schedule the job. 
Following the filtering phase, we enter the \textit{Ranking} phase where each node is given a score, and the node with the highest score is where the job is scheduled (as described in section \ref{QRIO Meta Server}. 
We support this process by implementing a custom ranking plugin. The ranking plugin contacts the QRIO Meta Server for the score of a certain job against a particular node (i.e. a backend). It repeats this process for all filtered nodes and then the appropriate node is selected based on the score. 

\section{Evaluation} 

In this section, we evaluate the performance of our infrastructure QRIO on some quantum computing benchmarks. The primary goal of our evaluation is to understand the efficacy of the different components of QRIO. 

\subsection{Evaluation Setup} \label{Eval setup}

QRIO is currently evaluated on a laptop with an Apple M2 chip and 8 GB of RAM. As a consequence, we cannot set up a Kubernetes cluster locally that can scale up to 100s of nodes. So, we simulate the core scheduler plugin's code locally i.e. outside the Kubernetes infrastructure. To do so, we set up a Python script that is functionally equivalent to the scheduler's code. This script filters the devices based on the user's job requirements and contacts the QRIO meta server to get the scores on the devices. However, this change does not affect the functionality or runtime of QRIO as the code being evaluated is identical to the scheduler's code. Our evaluation is identical to one where a 100 node Kubernetes cluster can be set up in the manner discussed in section \ref{Kubernetes Cluster}. \par
The current testbed of quantum resources for evaluation comprises 100 simulated quantum computers created with varying edge connectivity and error rates. Concretely, we have a random coupling map and error rate generation algorithm to generate devices with 15, 20, 27, 35, 50, 60, 78, 85, 95, and 100 qubits, each with edge connectivity (probability of having an edge between two qubits) - 0.1, 0.15, 0.3, 0.45, 0.54, 0.67, 0.7, 0.78, 0.89, and 0.98. Conceptually, a device with a high edge connectivity like 0.98 has 4 qubit connections for almost all qubits (we limit ourselves to at most 4 connections), whereas a device with lower edge connectivity like 0.1 has fewer connections per qubit. \par
An essential part of our simulated quantum devices is the error rates of the qubits and the basis gates. Specifically, we are concerned with two-qubit and single-qubit error rates for each device. We choose them uniformly randomly between values 0.01 and 0.7 - while these are higher than typical, these are useful for simulating the effects of noise on smaller quantum circuits. Moreover, we set other properties like T1, T2, Readout error, and Readout length in a similar fashion. For the basis gates of each device, we choose the gates \{$u_1$, $u_2$, $u_3$, CX\}. As a summary of our setup, we have presented the controllable parameters and our values in Table \ref{Setup Table}. \par

\begin{table}[htbp] 
\centering
\begin{tabularx}{\columnwidth}{|>{\raggedright}X|>{\raggedright\arraybackslash}X|}
\hline
\textbf{Controllable backend parameters} & \textbf{Our Values} \\ \hline
Number of qubits & 5, 20, 27, 35, 50, 60, 78, 85, 95, 100 \\ \hline
2-qubit gate error rate & 0.01 - 0.7 \\ \hline
1-qubit gate error rate & 0.01 - 0.7 \\ \hline
Readout rate & 0.05, 0.15 \\ \hline
T1 & 500e3, 100e3 \(\mu s\) \\ \hline
T2 & 500e3, 100e3 \(\mu s\) \\ \hline
Readout Length & 30 \(\mathrm{ns}\) \\ \hline
Edge connects probabilities & 0.1, 0.15, 0.3, 0.45, 0.54, 0.67, 0.7, 0.78, 0.89, 0.98 \\ \hline
Basis gates & $u_1$, $u_2$, $u_3$, CX \\ \hline
\end{tabularx}
\caption{Controllable Backend Parameters and our Values}
\label{Setup Table}
\end{table}

\subsection{Device Selection from Default  Topologies}
In this experiment, we evaluate the performance of QRIO's scheduler ranking plugin on various default topologies against the 100 different backends. Specifically, we choose the following defaults:
\begin{itemize}
    \item  A grid topology of 4 qubits.
    \item  A line topology of 6 qubits.
    \item  A ring topology of 7 qubits.
    \item  A heavy square topology of 6 qubits.
    \item  A fully connected circuit of 6 qubits.
\end{itemize}

As a baseline, we choose a \textit{random scheduler}. This scheduler randomly picks up a device in the list of filtered devices (all 100 devices are considered for this experiment) and chooses that for scheduling the user's job. This scheduling strategy lacks the decision made by QRIO's scheduler where it compares the scores on different backends and chooses the device which achieved the best score. \par
\begin{figure}[htbp]
\centering
\includegraphics[width=\columnwidth]{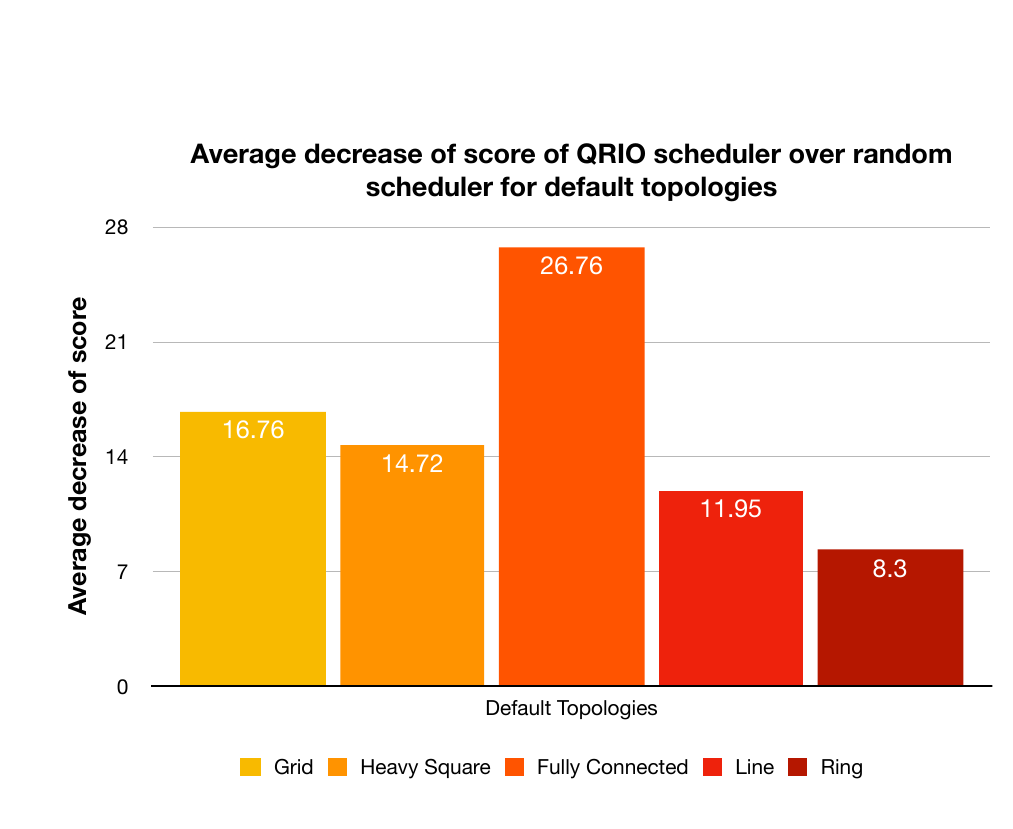}
\caption{Average decrease in score for QRIO scheduler compared to random scheduler} 
\label{fig:Evaluation 1}
\end{figure}

For this experiment, we compare the scores achieved by QRIO's scheduler and the scores achieved by the random scheduler. For the fairness of the evaluation, we repeated this experiment 25 times. We report the average \textbf{decrease} in scores of QRIO's scheduler compared to the random scheduler in \ref{fig:Evaluation 1} (As described in section \ref{Kubernetes Scheduler} it is always better to get a lower score and in our experiments, QRIO's scheduler always chooses a device which gets a score lower than the device which is chosen by the random scheduler). \par
As part of our results, we see that our simulated scheduler code works always better than the random choice. The differences in the amount of decrease can be explained by the underlying topologies being used. For instance, for the fully connected setting only a handful of devices can satisfy the topology request of the user. Specifically, these are devices with high qubit connectivity. In our cluster setup, we have almost 10 devices of this nature (These 10 devices are the devices that have an edge probability of 0.9). However, on average the random scheduling algorithm returns a device uniformly at random from the 100 devices and is less likely to be in these niche 10 devices more or less suitable for the fully connected default. \par
As for the ring default, all it requires is to have the qubits form a circle. In most devices, since no qubit is isolated, this property is satisfied with an appreciable confidence. So, the random scheduling algorithm does only 8.3\% worse than the simulated scheduler code.

\subsection{Fidelity Achieved by the Scheduler}
In this section, we evaluate the performance of QRIO for the fidelity ranking aspect of the scheduler. \par
As a setup for the evaluation, we choose the following algorithms
\begin{itemize}
    \item \textit{Random scheduler}: which chooses a random device from the 100 devices available.
    \item QRIO simulated scheduler.
    \item \textit{Oracle algorithm}: This algorithm scores the backends directly on the user's submitted circuit and does not convert it to a clifford circuit.
\end{itemize}

The most intriguing algorithm of the three is the last algorithm - \textit{Oracle algorithm}. Naturally, it is not possible to know the correct answer in this case and thus the fidelity cannot be defined. However, we run these circuits in a noise-free simulator, then record the correct outputs, and then compare them to calculate the fidelity. Since this algorithm returns the best possible device in the cluster which has the high fidelity it can give for the circuit, we call it the \textit{oracle algorithm}. To implement this in the real scheduler, we already require to know the right answer for the circuit which is not available at the time of scheduling. \par
\begin{figure}[htbp]
\centering
\includegraphics[width=\columnwidth]{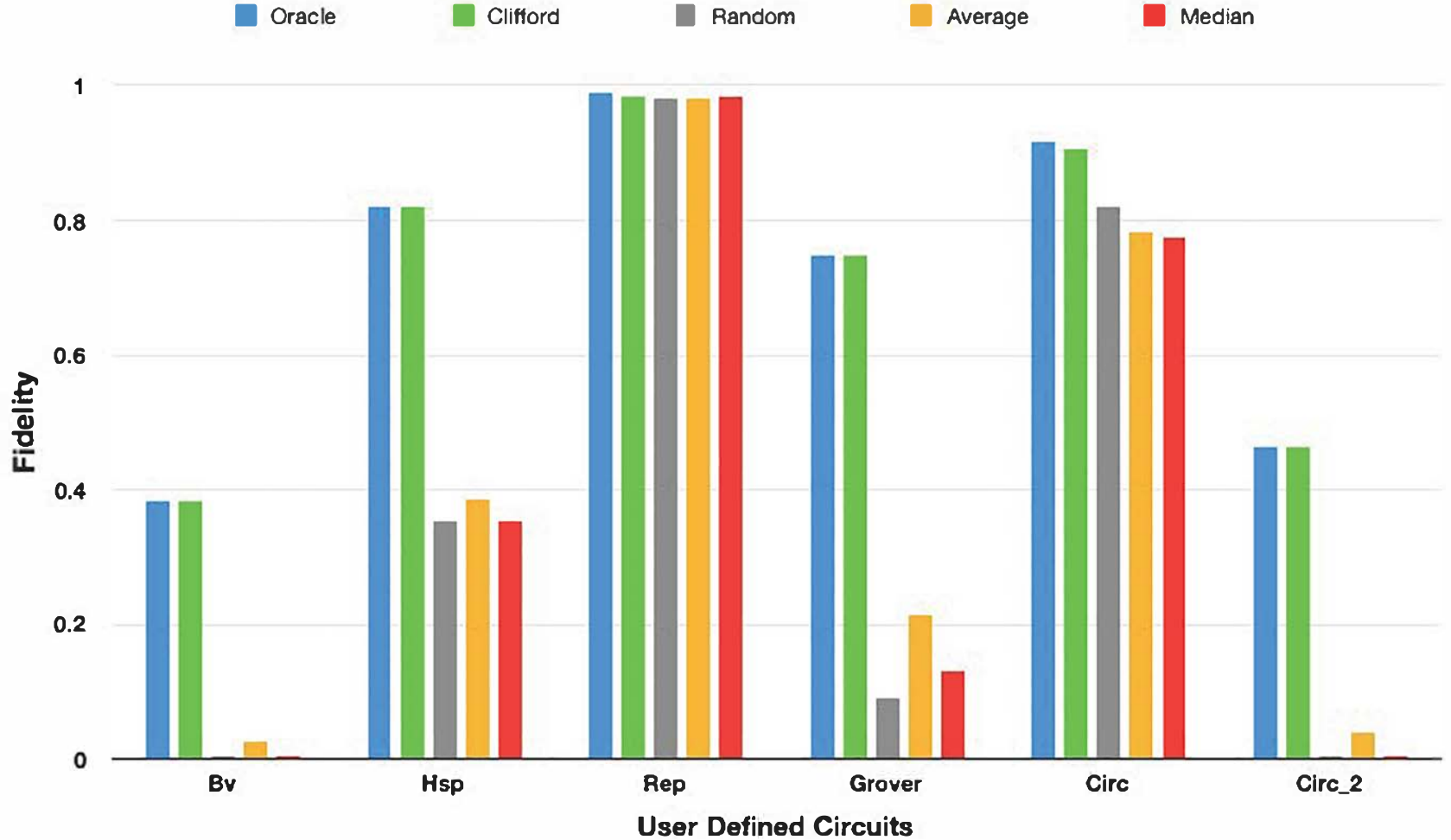}
\caption{Achieved Fidelity for user-defined circuit for Clifford simulation, Oracle simulation, Random simulation, Average and Median Fidelity} 
\label{fig:Evaluation 2}
\end{figure}

Throughout this experiment we consider the following circuits with a user demand of 100\% fidelity: \textit{Bernstien Vazirani(bv)\cite{bv}} (10 qubits), \textit{Hidden Subgroup Problem(Hsp)} (4 qubits), \textit{Grover Search(Grover)\cite{Grover96afast}} (3 qubits), \textit{Repetition Code Encoder(Rep)} (5 qubit), \textit{Circ} (random 7 qubit circuit), \textit{Circ\_2} (random 8 qubit circuit with 12 CX gates).
To perform this experiment, we first ran each of these algorithms to spit out the best device for each of the circuits. We then report the achieved fidelities for each of the circuits for the reported backends by the algorithms in figure \ref{fig:Evaluation 2}. \par
There are three aspects to the analysis of the results. Firstly, for the circuits \textit{bv, Hsp, Rep, Grover, Circ, Circ\_2} we cannot guarantee the user full 100\%. This is simply not possible for these circuits in the devices available in our cluster. However, in this case, we return the user the best possible device, that is close to the user's desired fidelity. \par
The second aspect of the analysis is the actual fidelities achieved by the three algorithms. Generally, the oracle always performs better than the simulated scheduler code. In the cases of all circuits except \textit{Circ} the clifford and oracle answers are very identical. This behavior is expected as the oracle's choice is always the best possible choice and since these circuits are not being changed in the clifford transformation (as they are already clifford), the fidelities are identical. In the case of \textit{Circ} the clifford fidelity is slightly lower than the oracle fidelity as we are transforming the circuit to a clifford circuit to get the right answer. Having said that, the fidelity is not very different from the oracle fidelity. This proves the effectiveness of our scheduler. We can allocate the best possible device given the fidelity choice of the user. \par
The device chosen by the Clifford method in our scheduler has better fidelity than the average fidelity and median fidelity of the devices in our test cluster. Moreover, this is an appreciable improvement in all of the test circuits except for \textit{Rep} as shown in \ref{fig:Evaluation 2}. This implies that the non-triviality of our scheduler is justified and brings benefits like - enhanced task outcomes and minimized low-fidelity risks.

\subsection{Device Choice based on User Requested Topology}
In this section, we present a qualitative analysis of our scheduler where the user provides a topology that they desire for their job. The primary aim of this evaluation is to show that QRIO respects the user's choice of topology and provides the one that is closest to it. \par
\begin{figure}[h]
\centering
\begin{subfigure}[t]{0.2\textwidth}
    \centering
    \fcolorbox{red}{white}{    \includegraphics[width=\columnwidth]{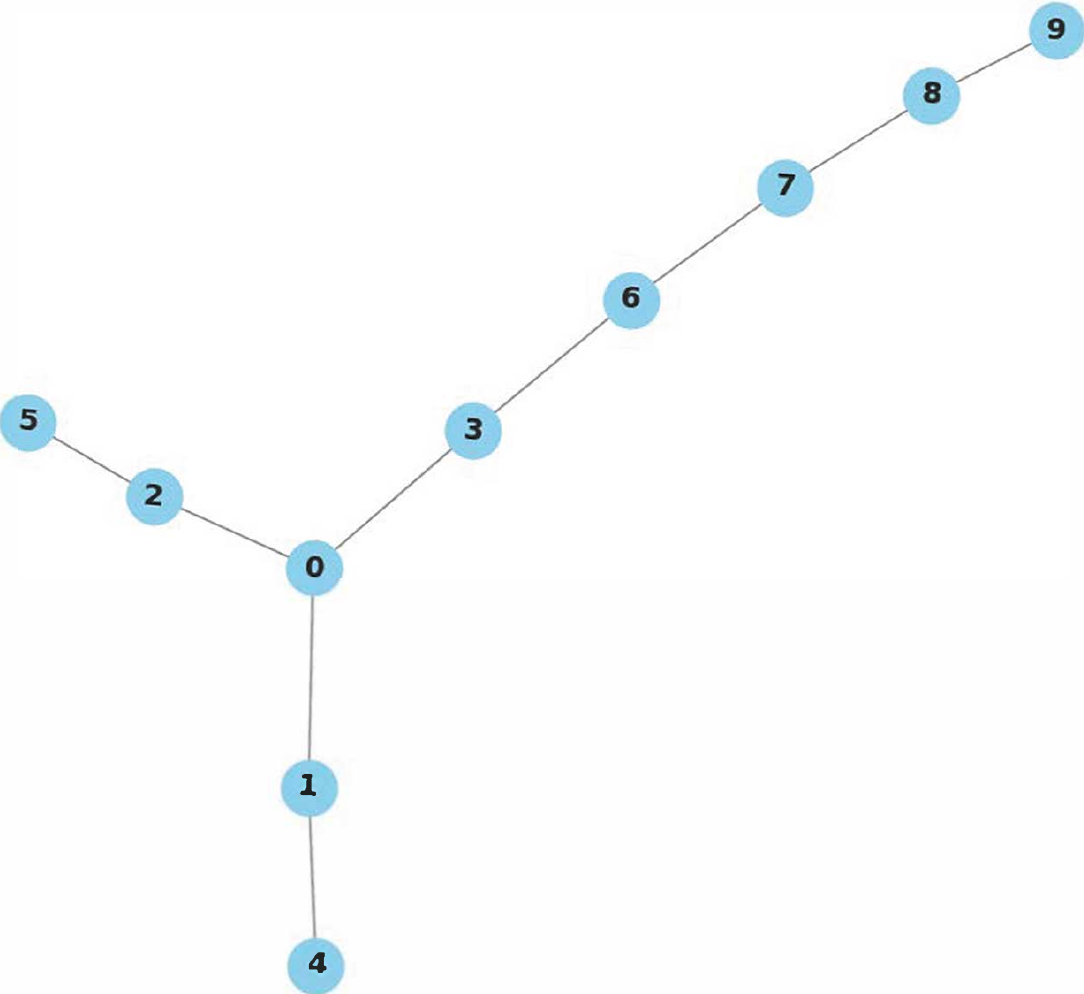}
}
    \label{fig:Evaluation_3_Device_avail_1}
\end{subfigure}

\begin{subfigure}[t]{0.2\textwidth}
    \centering
    \includegraphics[width=\columnwidth]{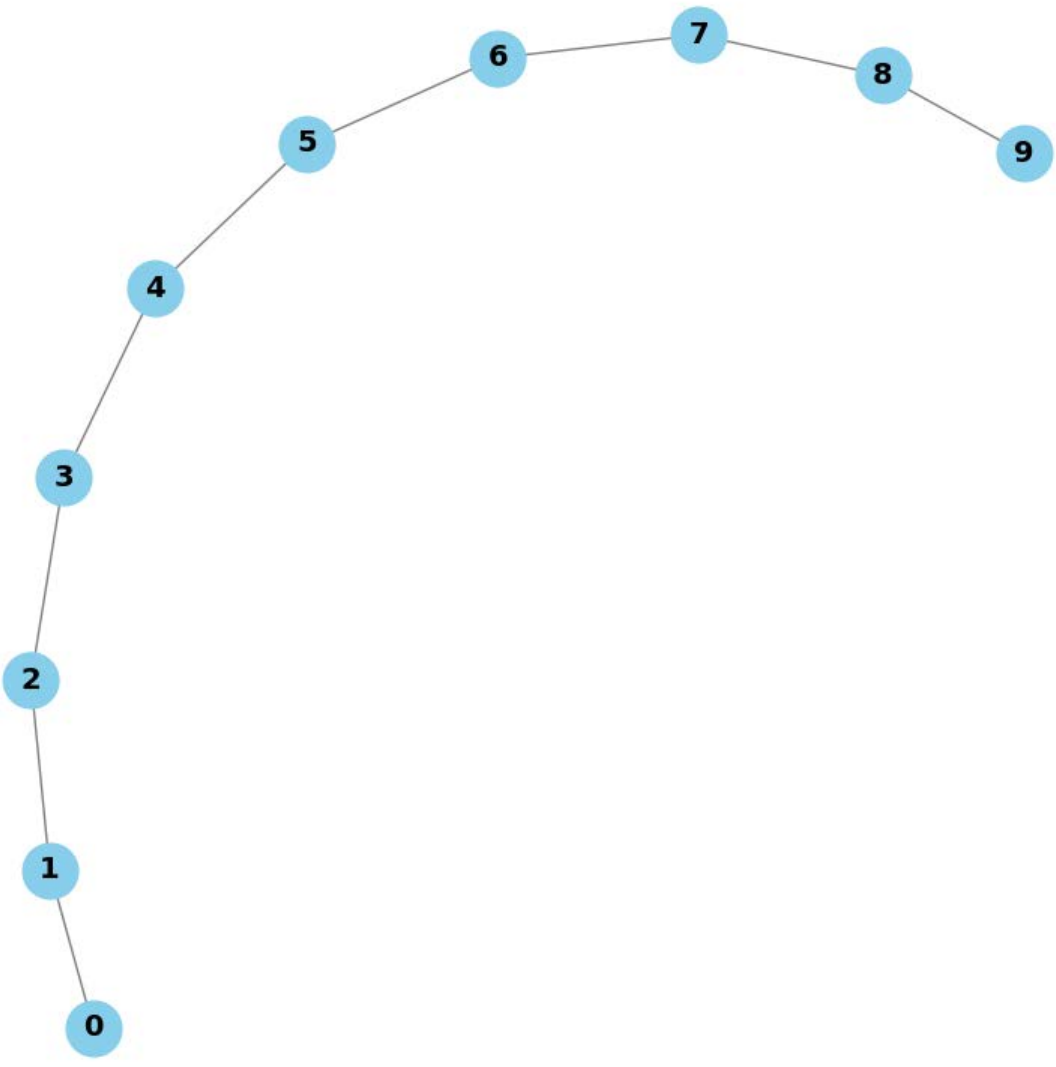}
    \label{fig:Evaluation_3_Device_avail_2}
\end{subfigure}
\begin{subfigure}[t]{0.2\textwidth}
    \centering
    \includegraphics[width=\columnwidth]{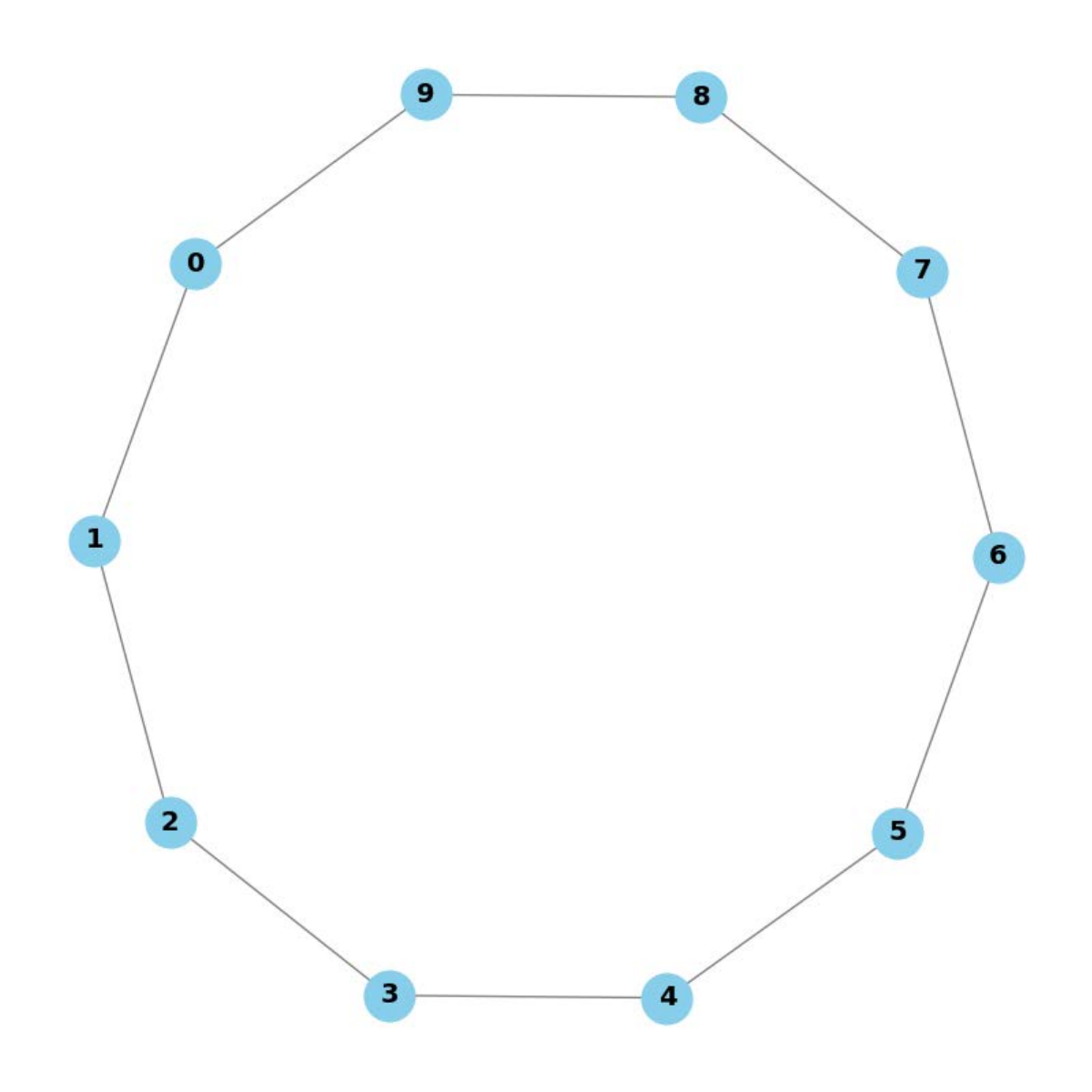}
    \label{fig:Evaluation_3_Device_avail_3}
\end{subfigure}
\caption{Available devices for the user-topology use case and chosen device boxed in red}
\label{fig:Evaluation_3_Devices}
\end{figure}

To get a visually comprehensible result, we created three devices of 10 qubits, with different topologies - A tree-like, A ring, and A line, shown in figure \ref{fig:Evaluation_3_Devices}. Since the aim of this experiment is to show the similarity of the topology chosen by the scheduler to the user's choice, we ignore factors like individual gate errors and qubits characteristics like T1, T2 times, etc. Specifically, we these values to be similar for all three devices. \par
\begin{figure}[htbp]
\centering
\includegraphics[width=\columnwidth]{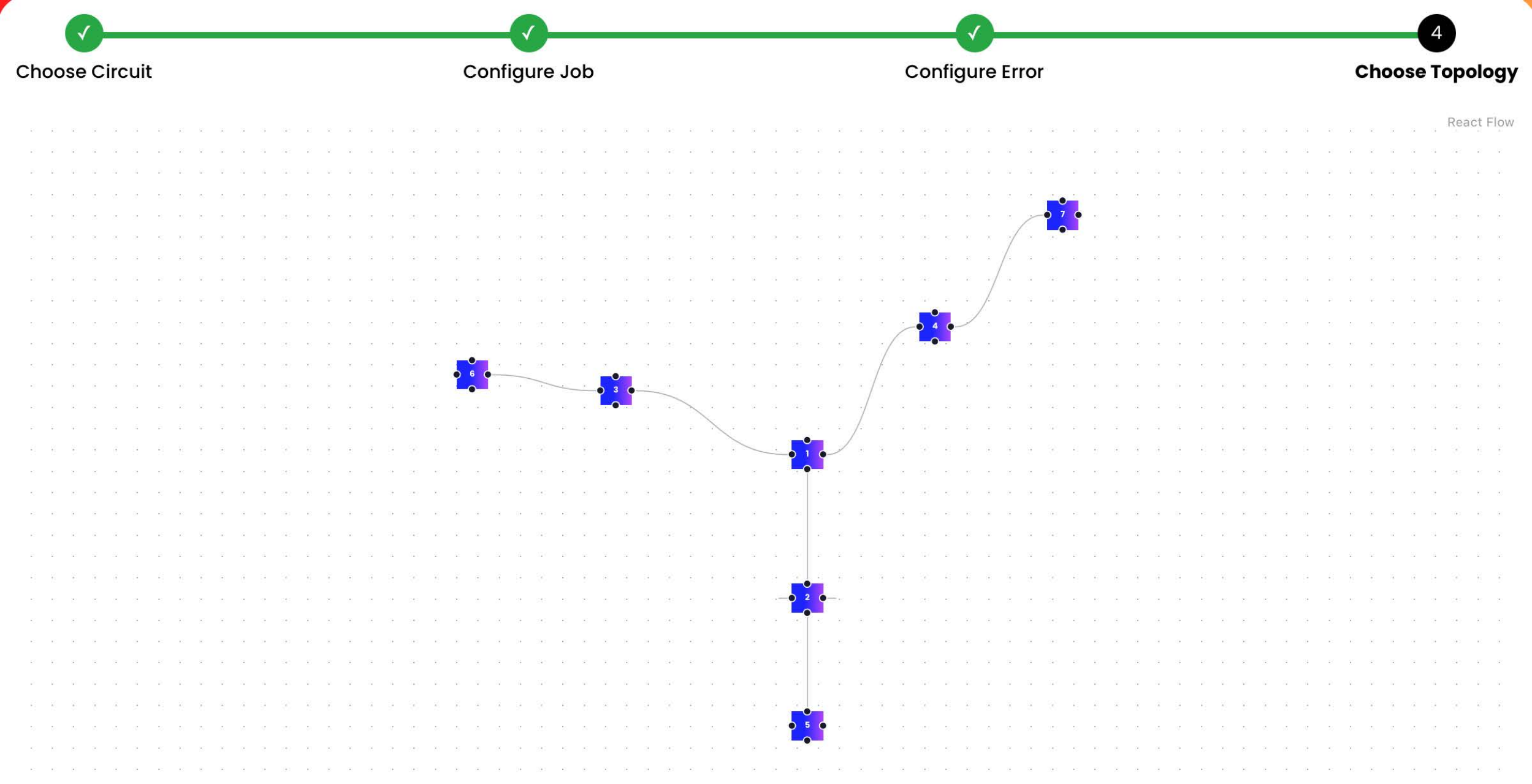}
\caption{User supplied topology} 
\label{fig:Evaluation 3_Topology}
\end{figure}

As for the user's topology, we create the topology using the topology drawing feature of QRIO as shown in figure \ref{fig:Evaluation 3_Topology}. We create a topology similar to the first device in \ref{fig:Evaluation_3_Devices} such that the final results are visually comprehensible. \par
The final device selected by the scheduler is shown in figure \ref{fig:Evaluation_3_Devices} boxed in red. As per our expectations, this is the closest to the user's choice of topology. The reason for this perfect similarity is due to the underlying graph subisomorphism feature of \textit{mapomatic}. Additionally, to ensure the fairness of the experiment, we repeated this 50 times and got the same result in all the runs.

\subsection{Filtered Devices from User Requested Characteristics}

In this section, we attempt to evaluate the filtering capabilities of the QRIO scheduler. The larger implication of this experiment is that we can save the computational resources of our scheduler. Therefore, the tighter the constraints, the lesser the number of devices the scheduler has to rank, thereby saving computational resources. \par
For this experiment, we use the original 100 backends created in the setup (section \ref{Eval setup}). Out of the controllable filtering parameters, the user changes the average two-qubit error rate parameter. Semantically, this parameter signifies the maximum two-qubit error rate the user can tolerate. As described in table \ref{Setup Table}, the user-controllable parameter also includes, T1, T2, and readout rates. However, this experiment is designed to showcase the efficacy and working of the filtering algorithm. The results will be similar for the other parameters. Moreover, for most quantum circuits and applications, the average two-qubit error rates are the factors that play a significant role. \par
\begin{figure}[htbp]
\centering
\includegraphics[width=\columnwidth]{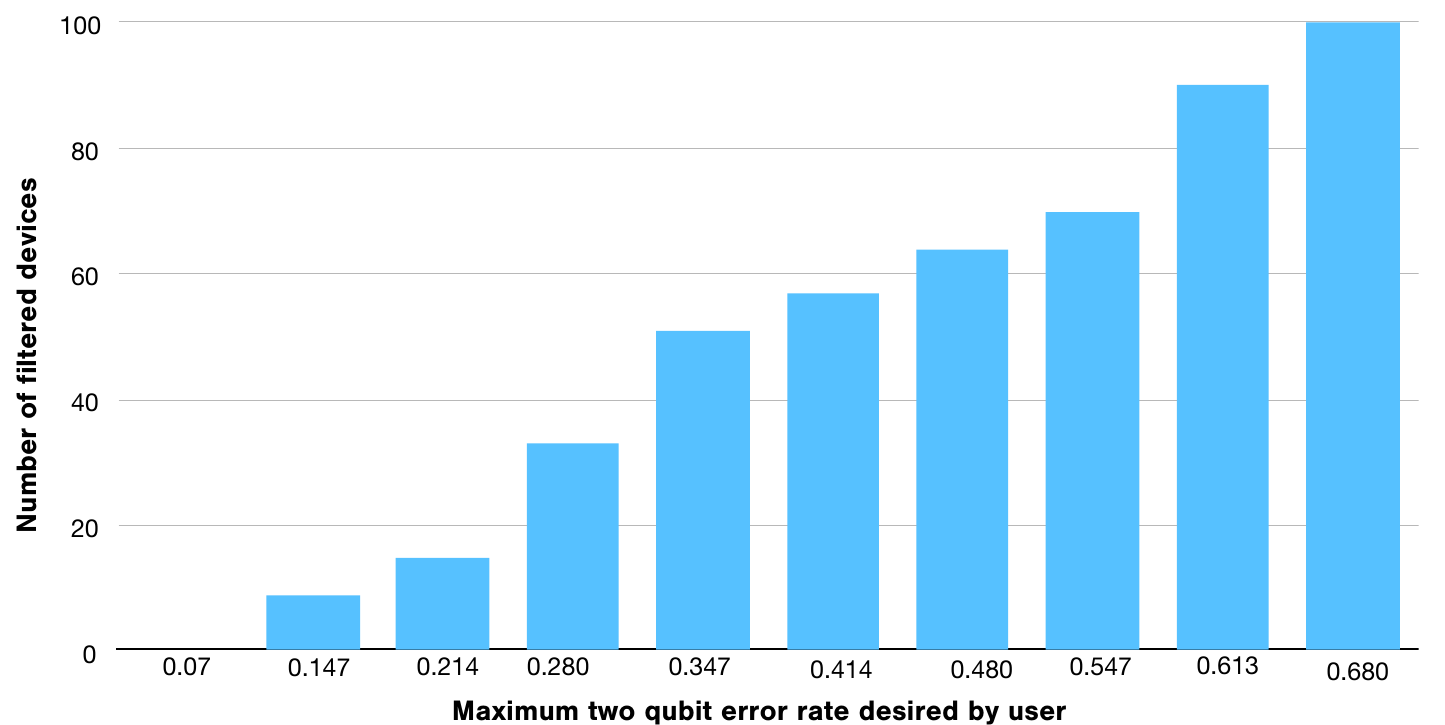}
\caption{Number of filtered devices for user-requested two-qubit gate errors } 
\label{fig:Evaluation 4_Filters}
\end{figure}

Our results are showcased in figure \ref{fig:Evaluation 4_Filters}. For two-qubit error rates as low as 0.07, we currently do not have devices in the cluster. So, the filtering algorithm returns zero devices. In the context of the application, such a situation would simply mean that the user's job is not fit for scheduling in the cluster. For high error rates such as 0.68, we get the entirety of the cluster. This is expected, as all devices in the cluster have a 2-qubit error rate $<=0.7$.

\section{Discussion and Future Work}
This paper presents an advancement towards the future of quantum cloud computing with efficient management of resources. It has certain shortcomings which we will develop in our future work.
(1) The current architecture of QRIO is highly geared towards the user i.e. the users have the highest amount of flexibility. However, there is limited flexibility in terms of the vendor. Although vendors can set up their custom architectures through the flexibility of Kubernetes, they currently cannot leverage a dashboard to do so; (2) Moreover, vendors have to specify their architectures as Backend instances of the Qiskit library. This might be a limitation to certain use cases, where vendors can't define the devices in terms of Qiskit Backends; (3) QRIO uses \textit{mapomatic} as its underlying library for scoring a topology sent by the user. This, however, incurs the disadvantages inherent in \textit{mapomatic}. For densely connected devices with almost full edge connectivity between qubits, it takes up to 45 minutes for the scoring process to return an answer. The problem is exacerbated when the user selects a topology that is densely connected and has a qubit count greater than 12-15 qubits; (4) The current infrastructure of QRIO is capable of handling one job scheduling request at a time. As part of our future work, we hope to scale this architecture to have the capability of handling multiple jobs at a time. This extension would require us to implement a job queue and also scale the architecture to run multiple physical nodes rather than one physical node.
\section{Conclusion}
In this paper, we describe QRIO a tool capable of automating the scheduling process for a Quantum Job provided by a user. QRIO appeals to a wide variety of audiences from as inexperienced as just supplying a fidelity number for their job to as experienced as supplying a topology suitable for their quantum job. Moreover, QRIO as a tool simplifies the user's lifecycle of scheduling their quantum job, by providing a GUI application and abstracting the underlying components responsible for scheduling. In future iterations, we aim to learn from the lessons we received from this work and build schedulers that match the level of sophistication current HPC schedulers have.


\bibliographystyle{IEEEtranS}
\bibliography{refs}
\end{document}